\documentclass[prb,aps,tightenlines,twocolumn,
showpacs,floatfix]{revtex4}
\usepackage{graphicx}
\usepackage{amssymb}
\usepackage{amsmath}
\usepackage{bm}
\usepackage{colordvi}
\usepackage{mathrsfs}

\begin{document}
\title{Single-parameter quantum charge and spin pumping in armchair graphene
  nanoribbons}
\author{Y. Zhou}
\author{M. W. Wu}
\thanks{Author to whom correspondence should be addressed}
\email{mwwu@ustc.edu.cn.}
\affiliation{Hefei National Laboratory for Physical Sciences at
  Microscale and Department of Physics, University of Science and
  Technology of China, Hefei, Anhui, 230026, China}

\date{\today}
\begin{abstract}
  We investigate quantum charge and spin pumping in armchair graphene
  nanoribbons under a single ac gate voltage connected with
  nonmagnetic/ferromagnetic leads via the nonequilibrium Green's function
  method. 
  In the case of nonmagnetic leads, where only part of the
  nanoribbon is subject to an ac gate voltage to break the left-right spatial
  symmetry, we discover that peaks of the charge pumping current appear at the
  Fermi energies around the subband edges in the ac-field-free region of
  the nanoribbon. 
  In the case of ferromagnetic leads with the lead magnetizations
  being antiparallel to break the left-right symmetry,
  similar peaks appear in the spin pumping current when the Fermi
  energies are around the edges of the the majority-spin subbands in the
  ferromagnetic leads. 
  All these peaks originate from the pronounced symmetry breaking in the
  transmissions with energies being around the corresponding subband edges.
  Moreover, we predict a {\em pure} spin current in the case
  of ferromagnetic leads with the whole graphene nanoribbon under an ac gate
  voltage. 
  The ac-field-strength and -frequency dependences of the pumping current
  are also investigated with the underlying physics revealed.
\end{abstract}

\pacs{72.80.Vp, 73.23.Ad, 72.25.-b, 72.40.+w}


\maketitle

\section{Introduction}
Graphene and its lower dimensional cousins, graphene nanoribbons (GNRs) and
carbon nanotubes, exhibit abundant new physics and potential applications and
hence have attracted much interest in recent
years.\cite{Novoselov_04,Geim_rev_nat,Neto_rev_09,Beenakker_rev,
Chakraborty_rev,Peres_rev,Mucciolo_rev,Sarma_rev}
Among different works in this field, the effect of an ac field on the
electrical and optical properties in these materials is one of the main
focuses of attention. Many interesting phenomena have been reported, such as the 
photon-assisted transport,\cite{Trauzettel_ac,Torres_ac_trans} the photovoltaic
Hall effect,\cite{Oka_09} the dynamical Franz-Keldysh effect\cite{Zhou_THz} and
the quantum pumping.\cite{Leek_CNT,Schomerus_ad,Zhu_ad,Tiwari_ad,ZWu_multi_pump,
Kaur_multi_pump,Gu_zigzag,Schomerus_single_pump,Torres_single_pump,Tiwari_spin,
Grichuk_spin_zigzag,Shan_ad_spin,Bercioux_SOC_spin}

The quantum pumping, which is highly related to the ratchet
effect\cite{Julicher_ratchet,Linke_ratchet,Reimann_ratchet,
Hanggi_ratchet,Ivchenko_ratchet} and the photogalvanic
effect,\cite{Belinicher_1980,Ivchenko_05,Ganichev_06} 
describes the generation of a direct current 
at zero bias in a spatially asymmetric system under ac
fields.\cite{Thouless_83,Altshuler_99} 
The quantum pumping of charge and spin currents has been investigated
theoretically\cite{Thouless_83,Altshuler_99,Brouwer_98,
Levinson_00,Vavilov_01,Moskalets_02,Mares_04,Kashcheyevs_04,
Strass_05,Arrachea_nonad,Torres_05,Agarwal_07,
HGuo_spin,Blaauboer_spin,Das_spin,CLi_spin,Romeo_spin} 
and observed experimentally in semiconductor quantum
dots,\cite{Switkes_QD,Kaestner_QD,Watson_QD_spin} 
quantum wires\cite{Blumenthal_QWi,Kaestner_QWi,Fujiwara_QWi} and 
also in carbon nanotubes.\cite{Leek_CNT}
Recently, the quantum charge and spin pumping in GNRs has also aroused growing
attention.\cite{Schomerus_ad,Zhu_ad,Tiwari_ad,ZWu_multi_pump,Kaur_multi_pump,
Gu_zigzag,Schomerus_single_pump,Torres_single_pump,Tiwari_spin,
Grichuk_spin_zigzag,Shan_ad_spin,Bercioux_SOC_spin} 
Most of these studies focus on the pumping involving more than one
time-dependent parameters,\cite{Schomerus_ad,Zhu_ad,Tiwari_ad,ZWu_multi_pump,
Kaur_multi_pump,Tiwari_spin,Grichuk_spin_zigzag,Shan_ad_spin,Bercioux_SOC_spin}
partially because the single-parameter pumping is only 
possible beyond the adiabatic approximation and thus needs
more complex theoretical tool.\cite{Brouwer_98} 
Nevertheless, the single-parameter pumping is more favorable to the
application, since the reduction in number of necessary contacts 
makes the scalable and low-dissipative device more
promising.\cite{Torres_single_pump} 

So far, there are few investigations on the single-parameter pumping in the
GNR. Torres {\em et al.}\cite{Torres_single_pump} studied the charge
pumping in the ribbon of small size, where the behaviour is dominated by
the resonant tunneling. 
However, the single-parameter pumping in a typical one-dimensional GNR, i.e.,
with large length and small width, has not yet been investigated.
Moreover, to the best of our knowledge, there is no work on single-parameter
spin pumping in this system. 
The aim of this study is to fill these spaces.

In this paper, we present a detailed study of the single-parameter quantum
charge and spin pumping in armchair GNRs contacted with
nonmagnetic/ferromagnetic leads via the nonequilibrium Green's function
method.\cite{Haug_2008} We first address the case of nonmagnetic leads, 
in which only part of the GNR is subject to an ac gate voltage to break 
the left-right spatial symmetry. 
It is discovered that peaks of the negative (positive) charge pumping current
appear at the Fermi energies around the energy
maximums (minimums) of the subbands in the ac-field-free region of the GNR. 
Then we turn to the case of ferromagnetic leads with the lead magnetizations
being antiparallel in order to break the left-right symmetry.
It is shown that a pure spin current can be achieved when
the whole GNR is under an ac gate voltage. 
We also predict peaks in the negative spin pumping current at the Fermi
energies around the energy maximums of the majority-spin subbands in the 
ferromagnetic leads. 
In the appendix, we discuss the time-dependent ballistic magnetotransport
in armchair GNRs contacted with ferromagnetic electrodes and show that the results 
under the cutoff-energy approximation, artificially introduced by Ding 
{\em et al.},\cite{Berakdar_11} are qualitatively different from
those obtained from the exact calculations.

This paper is organized as follows. In Sec.~II, we set up the tight-binding
Hamiltonian and the formula of the pumping current. The numerical
results are presented in Sec.~III. Finally, we summarize in Sec.~IV.

\section{Model and Formalism}
\label{Model}
\begin{figure}[tbp]
 \begin{center}
    \includegraphics[width=8.5cm]{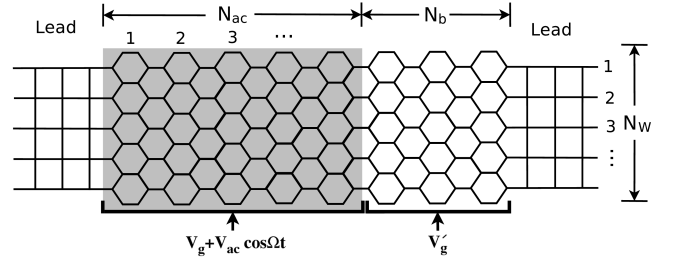}
  \end{center}
  \caption{Schematic view of the armchair GNR connected with two
    nonmagnetic/ferromagnetic leads. 
  }
  \label{structure_E}
\end{figure}

We consider an armchair GNR with a single gate voltage applied between two
nonmagnetic/ferromagnetic leads as shown in Fig.~\ref{structure_E}.
In this system, the tight-binding Hamiltonian can be written as 
\begin{equation}
  H=H_L+H_R+H_g+H_T,
\end{equation}
in which $H_L$ ($H_R$) represents the Hamiltonian of the left (right) lead,
$H_g$ stands for the Hamiltonian of the GNR and $H_T$ describes the hoping
between the GNR and the leads. These terms can be written as
\begin{eqnarray}
  {H}_{L}&=&\sum_{i_{L},\sigma} E_{L\sigma}
  c_{i_{L}\sigma}^\dagger c_{i_{L}\sigma}
  - \sum_{\langle i_{L},j_{L} \rangle, \sigma} t_{L\sigma}
  c_{i_{L}\sigma}^\dagger c_{j_{L}\sigma}, \\ 
  \nonumber
  {H}_{R}&=&\sum_{i_{R},\sigma} (E_{R0}+\sigma M_R \cos\theta)
  c_{i_{R}\sigma}^\dagger c_{i_{R}\sigma} \\
  && \hspace{-0.5cm} {}+ \sum_{i_{R},\sigma} M_R \sin\theta c_{i_{R}\sigma}^\dagger 
  c_{i_{R}{-\sigma}} - \hspace{-0.2cm}
  \sum_{\langle i_{R},j_{R} \rangle, \sigma} \hspace{-0.1cm} 
  t_{R\sigma} c_{i_{R}\sigma}^\dagger c_{j_{R}\sigma}, \\
  H_g&=&\sum_{i_g, \sigma} V_{i_g} d_{i_g\sigma}^\dagger
  d_{i_g\sigma} - \hspace{-0.2cm} \sum_{\langle i_g, j_g \rangle, \sigma} 
  \hspace{-0.1cm} t_g d_{i_g\sigma}^\dagger d_{j_g\sigma}, \\
  H_T&=&- \sum_{\alpha=L,R} \sum_{\langle i_g, j_\alpha \rangle, \sigma} 
  t_{T\sigma} d_{i_g\sigma}^\dagger c_{j_\alpha \sigma} + {\rm H.c.}.
\end{eqnarray}
Here $c_{i_{\alpha}\sigma}$ ($d_{i_g\sigma}$) and $c_{i_{\alpha}\sigma}^\dagger$
($d_{i_g\sigma}^\dagger$) are the annihilation and creation operators of the
electron with spin $\sigma$ on lattice site $i_{\alpha}$ ($i_g$) in the leads
(GNR); $\langle i,j \rangle$ denotes pair of nearest neighbors;
$E_{\alpha\sigma}=E_{\alpha 0}+\sigma M_{\alpha}$ 
are the on-site energy of the spin-up ($\sigma=+$) or spin-down ($\sigma=-$)
band in the ferromagnetic leads with
$E_{\alpha+}=E_{\alpha-}=E_{\alpha 0}$ 
for the nonmagnetic leads; $\theta$ is the angle between the
magnetization directions of right and left leads; 
$t_{\alpha\sigma}$ and $t_{T\sigma}$
represent the hoping parameters in the leads and between
the leads and the GNR, respectively, which are set to be equal to the hoping
parameter $t_g=2.7$~eV in the GNR, unless otherwise specified.
The on-site energy of the GNR
$V_{i_g}$ takes the values of $V_g + V_{ac}\cos(\Omega t)$ and 
$V_g^\prime$ in the ac-field-applied and -free regions of the GNR, 
respectively, with $V_{ac}$ and $\Omega$ being the magnitude and
frequency of the applied ac gate voltage.
Here we apply two static gate voltages $V_g$ and $V_g^\prime$ to modulate
the on-site energies in these two regions of the GNR independently to 
facilitate the identification of the influence of the pumping 
from different regions.

Exploiting the nonequilibrium Green's function method,\cite{Haug_2008} the
time-averaged current can be written as (the current flowing from left to right
is defined to be positive) 
\begin{eqnarray}
  \nonumber
  \overline{I}&=&\frac{e}{\hbar T_0}\int_0^{T_0}dt \int_{-\infty}^{\infty}
  \frac{d\varepsilon}{2\pi} \frac{d\varepsilon'}{2\pi} 
  \frac{d\varepsilon_1}{2\pi} e^{i(\varepsilon'-\varepsilon)t}\\ 
  \nonumber &&
  {} \times {\rm Tr}\left\{\hat{G}^a(\varepsilon,\varepsilon_1)
    \hat{\Gamma}_R(\varepsilon_1)
    \hat{G}^r(\varepsilon_1,\varepsilon') \hat{\Gamma}_L(\varepsilon')
    f_L(\varepsilon') \right. \\ && \left.
    {} - \hat{G}^a(\varepsilon,\varepsilon_1)
    \hat{\Gamma}_L(\varepsilon_1)
    \hat{G}^r(\varepsilon_1,\varepsilon') \hat{\Gamma}_R(\varepsilon')
    f_R(\varepsilon') \right\}.
  \label{I_ave}
\end{eqnarray}
Here $T_0=2\pi/\Omega$; $G^{r(a,<)}(\varepsilon,\varepsilon')$ 
is the retarded (advanced, lesser) Green's function in the ac-field-applied region of
the GNR; $f_\alpha(\varepsilon)$ is the Fermi distribution; 
$\hat{\Gamma}_{\alpha}(\varepsilon)=i[\hat{\Sigma}^r_{\alpha}(\varepsilon) 
-\hat{\Sigma}^a_{\alpha}(\varepsilon)]$, with
$\Sigma^{r(a)}_L(\varepsilon)$ and $\Sigma^{r(a)}_R(\varepsilon)$ representing the 
retarded (advanced) self-energies from the left lead and both the right
lead and the ac-field-free region of the GNR, respectively.
The self-energy has the form
\begin{equation}
  \hat{\Sigma}^\square_\alpha=
  \hat{R}_\alpha \begin{pmatrix} 
    \hat{\Sigma}^\square_{\alpha+} & 0 \\
    0 & \hat{\Sigma}^\square_{\alpha-}
  \end{pmatrix} \hat{R}^\dagger_\alpha,
\end{equation} 
where $\square=r,a$, the rotational matrix $\hat{R}_\alpha$ is defined as
\begin{equation}
  \hat{R}_\alpha=\begin{pmatrix} 
    \cos\frac{\theta_\alpha}{2} && -\sin\frac{\theta_\alpha}{2} \\
    \sin\frac{\theta_\alpha}{2} && \cos\frac{\theta_\alpha}{2}
  \end{pmatrix}
  \label{spin_rotation}
\end{equation} 
with $\theta_L=0$ and $\theta_R=\theta$.

Then we need to calculate $\hat{G}^{r}(\varepsilon,\varepsilon')$.
It is known that the eigenstates of the isolated GNR with and without an ac
gate voltage, labelled as ${\Phi}(t)$ and ${\Phi}_{0}(t)$
respectively, satisfy the relation\cite{Hanggi_98,Hanggi_05}
\begin{equation}
  {\Phi}(t)={\Phi}_{0}(t) 
  \exp\left\{-\frac{i}{\hbar}\int_{0}^t V_{ac}\cos(\Omega t) d\tau \right\}.
\end{equation}
Thus the Green's Function of the isolated ac-field-applied GNR 
$\hat{g}^r(\varepsilon,\varepsilon')$ has the form
\begin{eqnarray}
  &&\hat{g}^r(\varepsilon_r+n\Omega,\varepsilon_r'+m\Omega)=2\pi
  \delta(\varepsilon_r-\varepsilon_r') \overline{g}^r(\varepsilon_r,n,m),\\ 
  \nonumber
  &&\overline{g}^r(\varepsilon_r,n,m)= \sum_N J_{n-N}(\frac{V_{ac}}{\Omega})
  J_{m-N}(\frac{V_{ac}}{\Omega}) \hat{g}^r_0(\varepsilon_r+N\Omega). \\
  \label{G_0_2}
\end{eqnarray}
In these equations $\varepsilon_r\in[-\frac{\Omega}{2},\frac{\Omega}{2})$, 
$\hat{g}^r_0(\varepsilon)$ is the corresponding Green's function of this GNR
without the ac gate voltage, which can be obtained by the recursive
method.\cite{Datta_mesoscopic} From the Dyson equation 
\begin{equation}
  \hat{G}^r(\varepsilon,\varepsilon')=\hat{g}^r(\varepsilon,\varepsilon')
  + \int \frac{d\varepsilon_1}{2\pi} \hat{g}^r(\varepsilon,\varepsilon_1) 
  \hat{\Sigma}^r(\varepsilon_1) \hat{G}^r(\varepsilon_1,\varepsilon'),
\end{equation}
one obtains 
\begin{equation}
  \hat{G}^r(\varepsilon_r+n\Omega,\varepsilon_r'+m\Omega)=2\pi
  \delta(\varepsilon_r-\varepsilon_r') \overline{G}^r(\varepsilon_r,n,m),
  \label{G_form}
\end{equation}
where $\overline{G}^r(\varepsilon_r,n,m)$ is determined by
\begin{eqnarray}
  \nonumber
  \overline{G}^r(\varepsilon_r,n,m) &=& \overline{g}^r(\varepsilon_r,n,m)
  + \sum_{n_1} \overline{g}^r(\varepsilon_r,n,n_1) \\ && 
  {} \times \hat{\Sigma}^r(\varepsilon_r+n_1\Omega) 
  \overline{G}^r(\varepsilon_r,n_1,m).
\end{eqnarray} 
Substituting Eq.~(\ref{G_form}) into Eq.~(\ref{I_ave}), $\overline{I}$ can be
written as 
\begin{equation}
  \overline{I}=\frac{e}{h}\sum_{\sigma n} \int_{-\infty}^{\infty} 
  {d\varepsilon} [ T_{LR\sigma}^n(\varepsilon)f_L(\varepsilon) 
  -T_{RL\sigma}^n(\varepsilon)f_R(\varepsilon) ],
  \label{I2_ave}
\end{equation}
in which 
\begin{eqnarray}
  \nonumber
  T_{LR\sigma}^n(\varepsilon) 
  \nonumber
  &=& {\rm Tr}\left\{\hat{\Gamma}_{L\sigma}(\varepsilon)
    \overline{G}^a_\sigma(\varepsilon,\varepsilon+n\Omega)
    \hat{\Gamma}_{R\sigma}(\varepsilon+n\Omega)
  \right. \\ && \left. {}\times
    \overline{G}^r_\sigma(\varepsilon+n\Omega,\varepsilon)\right\}
  \label{T_n}
\end{eqnarray}
is the transmission probability from left to right leads involving
the absorption ($n>0$) or emission ($n<0$) of $|n|$ photons of electrons
with initial energy $\varepsilon$ and spin $\sigma$. 
Here we have limited ourselves in the parallel and
antiparallel configurations of the electrode magnetizations, and thus the
contribution from different spin bands can be calculated separately.

In the spatially asymmetric system, generally speaking 
$T_{LR\sigma}^n(\varepsilon) \ne T_{RL\sigma}^n(\varepsilon)$. 
Thus even when the external bias is absent, i.e.,
$f_L(\varepsilon)=f_R(\varepsilon)$, the time-average current can be nonzero
and a pump current emerges. In the zero temperature limit, the pumping current
with spin $\sigma$ is given by 
\begin{equation}
  {I}_{\rm pump}^{\sigma}=\frac{e}{h} \sum_{n} 
  \int_{-\infty}^{E_{\rm F}} {d\varepsilon}
  [ T_{LR\sigma}^n(\varepsilon)-T_{RL\sigma}^n(\varepsilon) ],
  \label{I_pump1}
\end{equation}
with $E_{\rm F}$ denotes the Fermi energy. The charge and spin pumping currents
are defined as $I_{\rm pump}^c={I}_{\rm  pump}^++{I}_{\rm pump}^-$ and 
$I_{\rm pump}^c={I}_{\rm  pump}^+-{I}_{\rm pump}^-$, respectively.
Further considering the time-reversal symmetry, one has
$T_{LR\sigma}^n(\varepsilon)=T_{RL\sigma}^{-n}(\varepsilon+n\Omega)$.
This means that the photon-assisted transmission with the initial and final
energies both below the Fermi energy is cancelled by the 
corresponding one in the opposite direction and hence cannot contribute to the
pump current. Consequently, ${I}_{\rm pump}^{\sigma}$ can be expressed as
\begin{equation}
  {I}_{\rm pump}^{\sigma}=\frac{e}{h} \sum_{n>0} 
  \int_{E_{\rm F}-n\Omega}^{E_{\rm F}} {d\varepsilon}
  [ T_{LR\sigma}^n(\varepsilon)-T_{RL\sigma}^n(\varepsilon) ].
  \label{I_pump2}
\end{equation}

\section{Numerical Results}
In this section, we present the numerical results of the quantum charge and spin
pumping currents in the armchair GNR connected with nonmagnetic/ferromagnetic 
leads. In our computation, the parameters are chosen to be $N_W=41$,
$N_{ac}=400$, $V_{ac}=\Omega=0.01 t_g$ and $V_g=V_g^\prime=0$, unless otherwise
specified. Since the width satisfies $N_W=3M+2$ with $M$ being an integer
number, thus this armchair GNR is metallic.\cite{Dresselhaus_ribbon} 
Also note that the length of the GNR is large enough to
totally suppress the contribution from the evanescent modes.\cite{Martin_07}

\begin{figure}[tbp]
 \begin{center}
    \includegraphics[width=6.8cm]{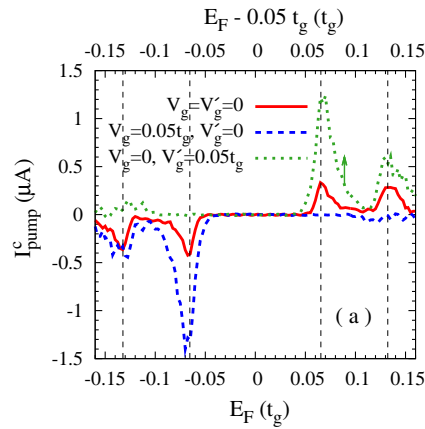}
    \includegraphics[width=5.5cm]{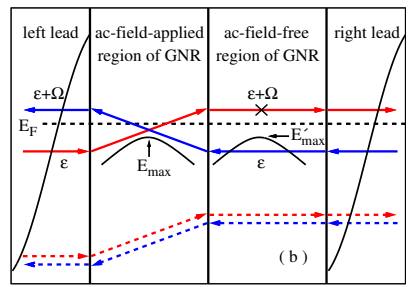}
  \end{center}
  \caption{(Color online) Armchair GNR connected with nonmagnetic leads.
    $N_{b}=100$.
    (a) Charge pumping current as function of the Fermi energy $E_{\rm F}$ for
    $V_g=V_g^\prime=0$ (red solid curve), $V_g=0.05 t_g$, $V_g^\prime=0$ (blue
    dashed curve) and $V_g=0$, $V_g^\prime=0.05 t_g$ (green dotted curve). Note
    that the scale of $E_{\rm F}$ for $V_g=0$ and $V_g^\prime=0.05 t_g$ is on
    the top of the frame. 
    The vertical black dashed lines indicate the energy maximums and
    minimums of the subbands in the ac-field-free region of the GNR.
    (b) The schematic illustration of one peak in the negative pumping current 
    for $V_g=V_g^\prime=0$. 
    The black dashed line indicates the Fermi energy.
    The red and blue solid arrows represent the transmissions 
    $T_{LR\sigma}^1(\varepsilon)$ and $T_{RL\sigma}^1(\varepsilon)$
    in the energy regime satisfying both 
    $E_{\rm F}-\Omega < \varepsilon < E_{\rm F}$ and 
    $E_{\rm max}^{\prime}-\Omega <\varepsilon < E_{\rm max}^{\prime}$. 
    The red and blue dashed arrows represent the transmissions
    $T_{LR\sigma}^1(\varepsilon)$ and $T_{RL\sigma}^{-1}(\varepsilon+\Omega)$ 
    for $\varepsilon<E_{\rm F}-\Omega$, which cancel each other.
  }
  \label{fig_current_ad}
\end{figure}

\subsection{Quantum pumping in armchair GNR  connected
with nonmagnetic leads} 

We first study the quantum charge pumping in the armchair GNR connected with
nonmagnetic leads, where the on-site energies are chosen to be
$E_{\alpha+}=E_{\alpha-}=0$.
We set a finite length of the ac-field-free region of the GNR, i.e., $N_b=100$, 
in order to break 
the symmetry between the transmissions $T_{LR\sigma}^n(\varepsilon)$ and 
$T_{RL\sigma}^n(\varepsilon)$ and hence induce the pumping
current.\cite{Schomerus_single_pump,Torres_single_pump}
In Fig.~\ref{fig_current_ad}(a), 
we plot the charge pumping current $I_{\rm pump}^c$ against the Fermi energy
$E_{\rm F}$ for $V_g=V_g^\prime=0$ (red solid curve). 
Here and hereafter, the energy zero point is chosen to be the 
Dirac point of the pristine GNR.
It is shown that peaks of negative (positive) current appear around the
energy maximums (minimums) of 
the subbands in the GNR indicated by the vertical black dashed lines.

These peaks are understood to originate from the pronounced symmetry breaking in
the transmissions with energies around the subband edges in the ac-field-free
region. 
We take one peak in the negative current as an example to illustrate this
physics in Fig.~\ref{fig_current_ad}(b).
Here we present the relevant subbands in the ac-field-applied and -free
regions of the GNR, whose energy maximums are labelled by $E_{\rm max}$ and 
$E_{\rm max}^{\prime}$, respectively.
The relevant subbands in the left and right leads are also shown, in which 
propagating modes exist in the whole energy range investigated here. 
The transmissions $T_{RL\sigma}^1(\varepsilon)$ and $T_{LR\sigma}^1(\varepsilon)$
with $E_{\rm max}^{\prime}-\Omega <\varepsilon < E_{\rm max}^{\prime}$ 
are plotted in this figure as blue and red solid arrows, respectively.
It is seen that the initial (final) energies of these transmissions
are lower (higher) than $E_{\rm max}^{\prime}$. 
Although both transmissions are allowed in the ac-field-applied region due to
the sideband effect, 
$T_{LR\sigma}^1(\varepsilon)$ is forbidden in the subband in the ac-field-free
region owing to the lack of the propagating modes above $E_{\rm max}^{\prime}$.
This makes the asymmetry in these transmissions become notable. 
It is also noted that only for $E_{\rm F}-\Omega < \varepsilon < E_{\rm F}$,
the single-photon assisted transmissions can
contribute to the pumping current as indicated by Eq.~(\ref{I_pump2}). 
As a result, a large negative pumping current appears when 
$E_{\rm max}^{\prime}-\Omega<E_{\rm F}<E_{\rm max}^{\prime}+\Omega$ and reaches
its peak value at the Fermi energy around $E_{\rm max}^{\prime}$.
Similar symmetry breaking can also be found in the multi-photon assisted
transmissions. After taking these transmissions into account, the peak of the 
pumping current appears at almost the same Fermi energy but with wider width. 
The presence of the peaks in the positive pumping current is based on the
similar physics. 
When the Fermi energy is around the minimum of one subband in the ac-field-free
region, this subband only contributes to the transmissions from
left to right, but not to the opposite ones. 
This symmetry breaking leads to the peak of positive current around this
minimum.

\begin{figure}[tbp]
  \begin{center}
    \includegraphics[width=6.cm]{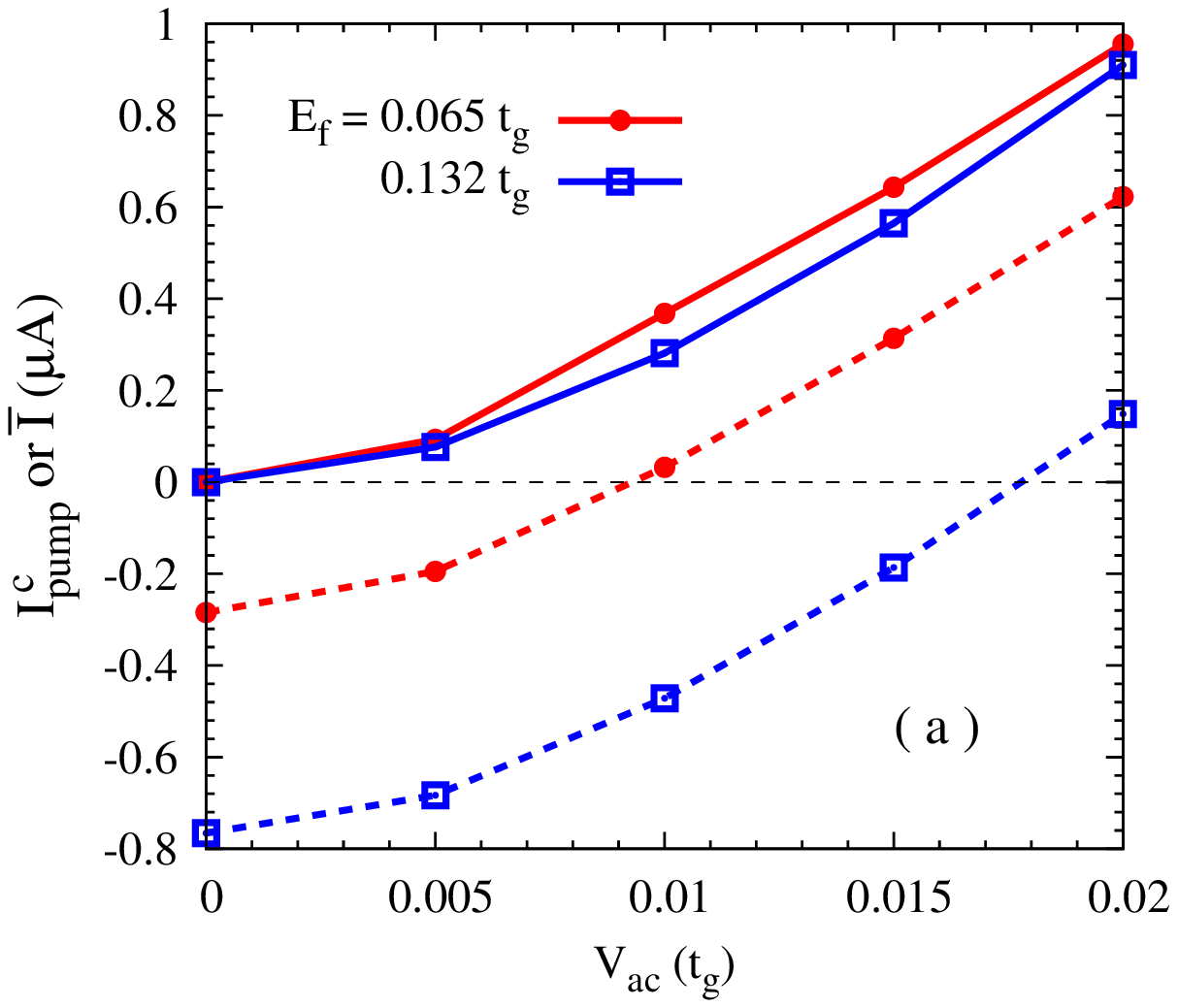}
    \includegraphics[width=6.cm]{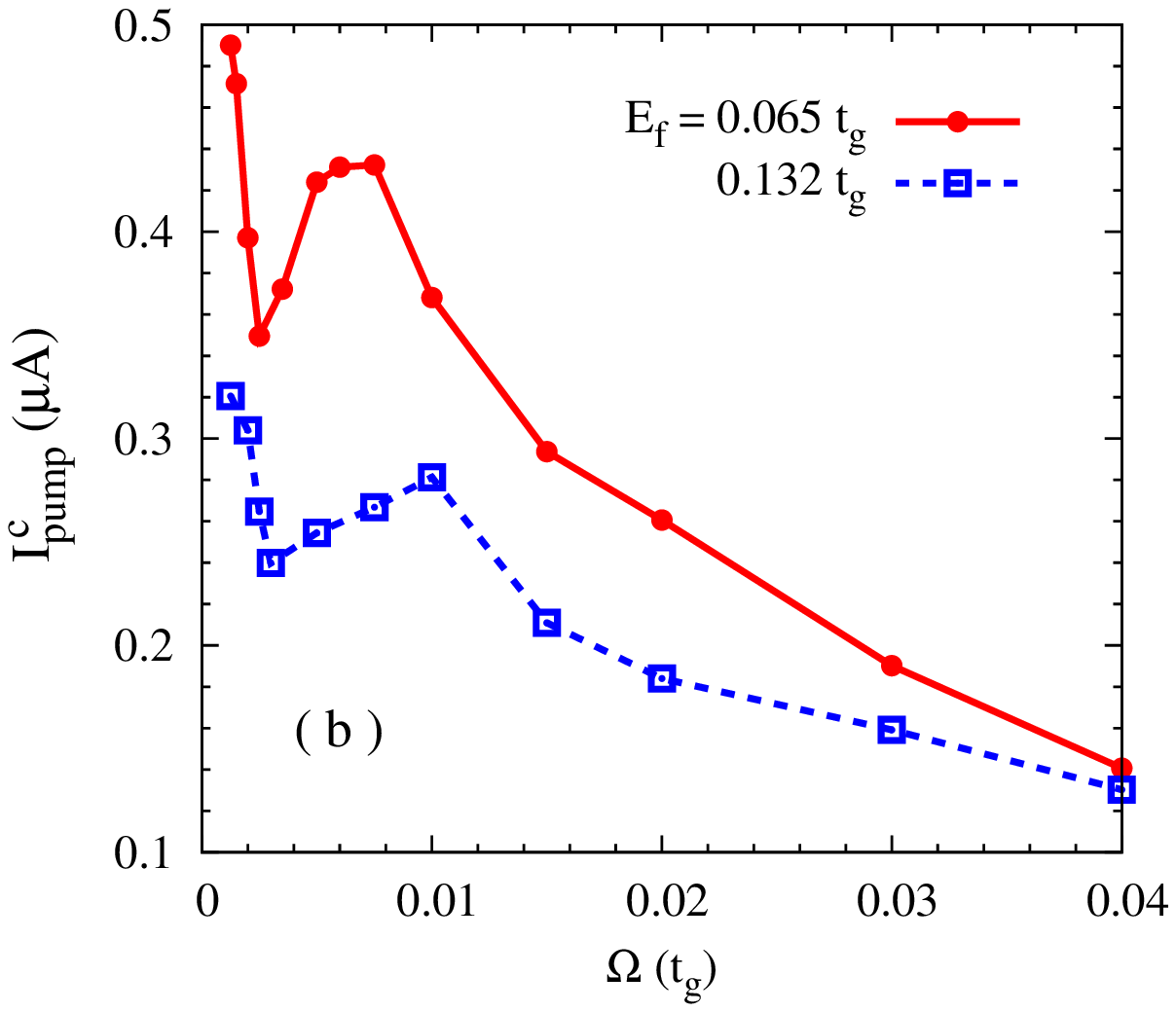}
  \end{center}
  \caption{(Color online) Armchair GNR connected with nonmagnetic leads.
    $N_{b}=100$. 
    (a)  Charge current as function of the ac-field strength $V_{ac}$
    with the frequency $\Omega=0.01 t_g$ for $E_{\rm F}=0.065 t_g$ (red 
    curves with $\bullet$) and $0.132 t_g$ (blue curves with $\square$).
    The solid and dashed curves represent the currents with no bias (i.e., the
    pumping current) and negative bias $E_{\rm F}^{L}=E_{\rm F}-\Omega/16$ and  
    $E_{\rm F}^{R}=E_{\rm F}+\Omega/16$.
    The black dashed line indicates the position of zero current.
    (b) Charge pumping current as function of the ac-field frequency
    $\Omega$ with $V_{ac}=0.01 t_g$ for $E_{\rm F}=0.065 t_g$ (red solid
    curve with $\bullet$) and $0.132 t_g$ (blue dashed curve with $\square$). 
  }
  \label{fig_current_dep}
\end{figure}

In order to further verify the above physics, 
we also plot $I_{\rm pump}^c$ with the modulated on-site energy
$V_g=0.05 t_g$, $V_g^\prime=0$ (blue dashed curve) and $V_g=0$, 
$V_g^\prime=0.05 t_g$ (green dotted curve) in Fig.~\ref{fig_current_ad}(a). 
It is seen that in the former case, the Fermi energies corresponding to peaks
of negative current are the same as those in the unmodulated case; whereas in
the latter one, the Fermi energies corresponding to peaks of positive current
are higher than the unmodulated ones by $0.05t_g$. 
These behaviours agree with the physics discussed above: 
it is the subband edges in the ac-field-free region that determine the energy
range where the pronounced symmetry breaking appears, and hence the Fermi
energies corresponding to the peaks in the pumping current. 
Moreover, the modulation of the on-site energy also brings about more
interesting phenomena. 
It is shown that in the case with $V_g=0.05 t_g$ and $V_g^\prime=0$ 
($V_g=0$ and $V_g^\prime=0.05 t_g$), the peaks of negative 
(positive) current appear with larger magnitude than those in the unmodulated
case, while all peaks of positive (negative) current disappear. 
These behaviours for $V_g=0.05 t_g$ and $V_g^\prime=0$ can be
understood as follows.
When the Fermi energy is around $E_{\rm max}^{\prime}$ shown in
Fig.~\ref{fig_current_ad}(b), the relevant transmissions from right to left in
this case are stronger than those without modulation, 
since the increase of $E_{\rm max}$ makes more states in 
the subband in the ac-field-applied region contribute to these
transmissions through the sideband effect. 
Consequently, the asymmetry in the transmissions becomes more pronounced and
thus the magnitude of the peak in the negative current becomes larger. 
On the other hand, when the Fermi energy is around the minimum of one
subband in the ac-field-free region, the relevant transmissions in both
directions are forbidden in the corresponding subband in the ac-field-applied
region, because the energy minimum of this subband is much higher than the
related energies of transmissions. 
This cancels the asymmetry in the transmissions and
hence suppresses the peak in the positive current. 
The behaviours in the case with $V_g=0$ and $V_g^\prime=0.05 t_g$ can be
understood in the similar way.

We stress that the above picture is only valid in the long ribbons, where the
energy spectrum can be considered quasi-continuous, but invalid in the short
ribbons. 
In the latter case, the field frequency is smaller than the energy level
spacing, thus the behaviour of the pumping current is dominated by the resonant
tunneling,\cite{Torres_single_pump} instead of the physics presented above. 

Then we turn to the ac-field-strength and frequency dependences of
the pumping current. 
In Fig.~\ref{fig_current_dep}(a), the pumping current is plotted against the
ac-field strength $V_{ac}$ with frequency $\Omega=0.01 t_g$ for two Fermi
energies $E_{\rm F}=0.065 t_g$ and $0.132 t_g$, which correspond to the two
peaks of positive current in Fig.~\ref{fig_current_ad}(a). 
It is seen that $I_{\rm pump}^c$ increases  
monotonically with $V_{ac}$ for various Fermi energies. 
This is because the photon-assisted transmissions are enhanced when the ac 
field becomes stronger. Besides the pumping currents, 
i.e., the current with no bias, the currents with negative bias 
$E_{\rm F}^{L}=E_{\rm F}-\Omega/16$ and $E_{\rm F}^{R}=E_{\rm F}+\Omega/16$ 
are also plotted in this figure by dashed curves. 
It is seen that the current is negative in the absence of an ac field. When
the ac field is strong enough, the pumping current can surpass the
current driven by the negative bias and hence make the net
current become positive, i.e., along the direction {\em opposite} to the
external bias. 

In Fig.~\ref{fig_current_dep}(b), we plot the pumping currents against the 
field frequency $\Omega$ with strength $V_{ac}=0.01 t_g$ for Fermi
energies $E_{\rm F}=0.065 t_g$ and $0.132 t_g$.
It is seen that $I_{\rm pump}^c$ presents a nonmonotonic frequency dependence
when $\Omega<V_{ac}$. The scenario is as follows.
On one hand, with an increase of the frequency, $V_{ac}/\Omega$ decreases and
thus the photon-assisted transmissions become weaker. On the other hand, the
energy range of the transmissions contributing to the pumping current increases
with $\Omega$ as shown in Eq.~(\ref{I_pump2}). The competition of these two
factors leads to the complex frequency dependence.
For $\Omega>V_{ac}$, the pumping current is shown to decrease with increasing
frequency. This is because the ac field becomes ineffective in exciting photons
when frequency is high enough.

\begin{figure}[tbp]
  \begin{center}
    \includegraphics[width=7.5cm]{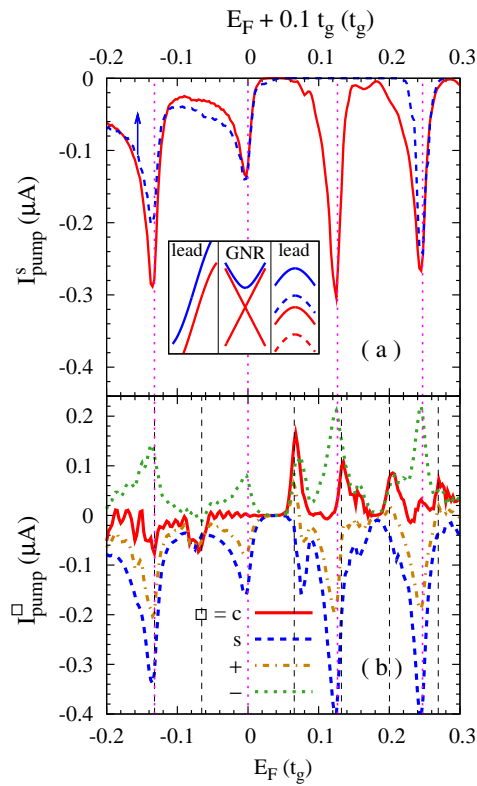}
  \end{center}
  \caption{(Color online) Armchair GNR connected with ferromagnetic leads 
    with the lead magnetizations being antiparallel.
    (a) Spin pumping current $I_{\rm pump}^s$ as function of the Fermi energy
    $E_{\rm F}$ for $E_{L+}=0$, $E_{R+}=-3 t_g$ 
    (red solid curve) and $-3.1 t_g$ (blue dashed curve). $N_{b}=0$. 
    Note that the scale of $E_{\rm F}$ for $E_{R+}=-3.1 t_g$ is on
    the top of the frame. In the inset, the spin-up
    subbands in the GNR and the ferromagnetic leads relevant to the peaks 
    of the pumping current at $E_{\rm F}=0$ (red curves) and $0.126 t_g$ (blue
    curves) are schematically plotted. 
    The solid curves represent the subbands in the case with
    $E_{L+}=0$ and $E_{R+}=-3 t_g$, while the dashed ones in the right panel
    represent the subbands in the right lead for $E_{R+}=-3.1 t_g$. 
    (b) Charge and spin pumping currents together with the spin-up and -down
    pumping currents against the Fermi energy $E_{\rm F}$ for $E_{L+}=0$ and
    $E_{R+}=-3 t_g$. $N_{b}=100$. 
    In these figures, the vertical black dashed (pink dotted) lines indicate the 
    energy edges of the subbands in the ac-field-free region of the GNR
    (the majority-spin subbands in the ferromagnetic leads).
  }
  \label{fig_spin_current}
\end{figure}

\subsection{Quantum pumping in armchair GNR connected 
with ferromagnetic leads} 
Now we investigate the quantum charge and spin pumping in the armchair GNR
connected with ferromagnetic leads. 
In order to break the left-right symmetry,
the magnetizations of the two leads are set to be antiparallel, i.e.,
$E_{L\sigma}=E_{R-\sigma}$, and hence the ac-field-free region of the GNR
becomes unessential. 
We first focus on the case without this region, i.e., $N_b=0$. 
In this case, the system satisfies the symmetry 
$R_{\pi} \mathcal{R}_{LR}$, where $R_{\pi}$ is the spin rotation operator
defined by Eq.~(\ref{spin_rotation}) with rotation angle $\pi$ 
and $\mathcal{R}_{LR}$ is the spatial reflection operator in the direction from
left to right. 
Thus, the spin-up and -down pumping currents always satisfy the relation 
$I_{\rm pump}^{+}=-I_{\rm pump}^{-}$. Consequently, the charge pumping
current vanishes and a {\em pure} spin current can be achieved.

We plot the spin pumping current $I_{\rm pump}^s$ for $E_{L+}=0$ and
$E_{R+}=-3t_g$ as red solid curve in Fig.~\ref{fig_spin_current}(a).
It is shown that peaks of negative current appear at the Fermi energies around
the maximums of the majority-spin (${L-}$ and ${R+}$) subbands in the
ferromagnetic leads, as indicated by the vertical pink dotted lines. 
In order to reveal the underlying physics, we schematically plot 
the spin-up subbands in the GNR and the ferromagnetic in the inset 
of Fig.~\ref{fig_spin_current}(a). 
Here only the subbands relevant to the peaks at $E_{\rm F}=0$ (red solid
curve) and $0.126 t_g$ (blue solid curve) are shown. 
Comparing this inset with Fig.~\ref{fig_current_ad}(b), 
one reaches the conclusion that
the physics of these peaks is similar to that in the case of nonmagnetic
leads: they come from the pronounced asymmetry in the transmissions with
energies around the maximums of the majority-spin subbands in the ferromagnetic
leads. 
In order to further verify this physics, we low the on-site energy $E_{R+}$
by $0.1 t_g$ and plot the corresponding $I_{\rm pump}^s$ 
as blue dashed curve in Fig.~\ref{fig_spin_current}(a). 
One observes that the Fermi energies corresponding to peaks also decrease by
$0.1 t_g$ as expected. 
The only exception is that 
the peak previously at $E_F=0.126 t_g$ is absent in this case. 
This feature is based on the similar physics leading to the absence of the
corresponding peaks in the case of nonmagnetic leads. 
As shown in the inset of Fig.~\ref{fig_spin_current}(a),
the energy gap between the subbands relevant to the absent peak in the GNR 
(blue solid curve in the middle panel) and right lead (blue dashed curve in the
right panel) is much larger than the photon energy.
Consequently, these subbands cannot contribute to the relevant transmissions in
either direction. This leads to the absence of the peak. 

Now we turn to the case with finite $N_b$. 
Due to the breaking of the symmetry $R_{\pi} \mathcal{R}_{LR}$, both charge and
spin pumping currents can be finite. 
In Fig.~\ref{fig_spin_current}(b), we plot these two pumping currents 
against the Fermi energy for $E_{L+}=0$ and $E_{R+}=-3 t_g$ as red solid and
blue dashed curves, respectively. 
By comparing these results with those in Figs.~\ref{fig_current_ad}(a) and
\ref{fig_spin_current}(a), similar features can be seen between the spin
or charge pumping currents here and in the case of ferromagnetic
leads for $N_b=0$ or the case of nonmagnetic leads, i.e.,
peaks in the spin (charge) pumping current appear at
the Fermi energies around the edges of the majority-spin subbands in the leads
(the subbands in the ac-field-free region of the GNR).
The scenario is as follows. 
Here the pumping current comes from two kinds of symmetry breaking.
The first is due to the antiparallel magnetizations of the ferromagnetic leads,
which induces peaks where the Fermi energies are around the edges of the
majority-spin subbands in the leads, as discussed above. 
From Fig.~\ref{fig_spin_current}(b), one further observes that
these peaks have similar magnitude but the opposite sign in the spin-up
(yellow chain curve) and -down (green dotted curve) pumping currents. 
This results in peaks in the spin pumping current but provides negligible
contributions in the charge pumping current. 
The second kind of symmetry breaking is from the presence of the
ac-field-free region of the GNR, which induces peaks where the Fermi energies
are around the subband edges in this region, as demonstrated in Sec.~IIIA.
These peaks tend to cancel each other in the spin pumping current, but
manifest themselves in the charge pumping current. 
Under the joint effect of these two kinds of symmetry breaking, the behaviour of
the spin or charge pumping current resembles that in the case of ferromagnetic
leads for $N_b=0$ or the case of nonmagnetic leads. 
Moreover, 
some new features are induced due to the interplay of these two kinds of
symmetry breaking, 
such as the peaks in the negative spin pumping current around $E_{\rm F}=0.075
t_g$ and the peak in the positive charge pumping current around $E_{\rm F}=0.247 
t_g$. 
In addition, we verify the ac-field-strength and frequency dependences of
the spin and charge pumping currents with ferromagnetic leads resemble those of
the charge pumping current with nonmagnetic leads and hence do not repeat here.

\section{Summary and discussion}
In summary, we have performed a detailed study of the single-parameter quantum
charge and spin pumping in the armchair GNR connected with
nonmagnetic/ferromagnetic leads via the nonequilibrium Green's function method. 
We first study the charge pumping in the case of nonmagnetic
leads. Here only part of the GNR is subject to an ac gate voltage in order to
break the left-right spatial symmetry. 
We discover that peaks of the negative (positive) charge pumping current appear
when the Fermi energies are around the energy maximums (minimums) of the
subbands in the ac-field-free region of the GNR. 
This phenomenon comes from the pronounced symmetry breaking in the
transmissions with energies being around these subband edges. 
We also discuss the pumping current with the modulated on-site energies in
ac-field-free and -applied regions of the GNR. 
Our results show that the Fermi energies corresponding to peaks are
influenced by the energy-level shifting in the ac-field-free region, but
independent of that in the ac-field-applied region. 
This is in agreement with the physics presented above. 
Furthermore, we find that the modulation of the on-site energies also induces
the absence of some peaks.
This is because the relevant transmissions are forbidden due to the lack of
the propagating modes in the corresponding subbands in the ac-field-applied
region. 

The ac-field-strength and -frequency dependences of
the pumping current are also investigated. We show that the pumping
current increases monotonically with the ac-field strength due to the enhanced
photon-assisted transmissions. 
Moreover, we discover that at low ac-field frequency, the pumping current exhibits a
nonmonotonic frequency dependence. This is due to the competition of the
weakened photon-assisted transmissions and the increasing energy space of the
states contributing to the pumping current.
At high frequency, the pumping current decreases monotonically with increasing
frequency, since the ac field becomes ineffective in exciting photons.

More interesting features are seen in the charge and spin pumping in the
case of ferromagnetic leads. 
Here we set the magnetizations of the two leads to be antiparallel
to break the left-right symmetry. 
In the case with the whole GNR under an ac field,
we show that the charge pumping current vanishes and a {\em pure}
spin current can be achieved.
We also find that peaks of negative spin pumping current appear at the Fermi 
energies around the maximums of the majority-spin subbands in the 
ferromagnetic leads. 
The physics is similar to the pumping peaks with nonmagnetic leads.
In the case with only part of the GNR subject to an ac gate
voltage, both the charge and spin pumping currents can be finite. 
Our calculations show that the behaviour of the spin pumping current is 
similar to that in the case of ferromagnetic leads with the whole GNR under an
ac field, while the behaviour of the charge pumping resemble that with
nonmagnetic leads.

Finally, we stress that the main features in the charge and spin pumping currents
predicted here do not depend on the specific properties of the leads.
Thus they appear not only in the case of the simple-squared-lattice leads used
in our model, but also in the case of the other types of leads, e.g., the
two-dimensional superlattice\cite{Andres_2DSL,Kleinerta_2DSL} and graphene leads.
This provides more choices for the experimental investigations.

\begin{acknowledgments}
  This work was supported by the National Basic Research Program of China under
  Grant No.\ 2012CB922002 and the National Natural Science Foundation of China
  under Grant No.\ 10725417. Y.Z. would like to thank P. Zhang and H. Tong for
  their critical reading of this manuscript. 
\end{acknowledgments}

\appendix*
\section{Magnetotransport in graphene under ac gate voltage}
\label{magnetotransport}
Recently, Ding {\em et al.}\cite{Berakdar_11} studied the magnetotransport of a
similar structure, i.e., graphene connected with ferromagnetic leads in the
presence of an ac gate voltage. 
Similar to our model, they treated the transport inside the graphene and leads as
the ballistic transport, i.e., without considering any scattering. 
However, their treatment of the interface between graphene and 
leads is very different.
They neglected the momentum dependence of the coupling matrix and replaced all
coupling matrix elements by only one 
phenomenological parameter. This leads to a divergence in the transmission.
To remove the divergence, they artificially introduced a cutoff energy.
It is exactly due to this cutoff energy, many pronounced features in the
time-dependent magnetotransport are predicted in that paper. 

In the following, we will demonstrate that these pronounced features are just
the artificial results induced by the inappropriate treatment of the interface
and the introduction of the cutoff energy. 
Since there is no cutoff energy in our model, we can calculate the transport
properties without any approximation used in Ref.~\onlinecite{Berakdar_11}, and
compare the obtained results with corresponding one with the cutoff energy. 
In the main text of this investigation, the parameters of the ferromagnetic leads
are set as $E_{L+}\ne E_{L-}$ and $t_{T\sigma}=t_g$.
In this appendix, in order to reveal the problems in that paper more clearly, we
change the parameters to $E_{L+}=E_{L-}=0$ and $t_{T\sigma}=t_g\sqrt{1+\sigma P}$
with $P=0.8$ to make $\hat{\Gamma}_{L+}/\hat{\Gamma}_{L-}=(1+ P)/(1- P)$ as
done in Ref.~\onlinecite{Berakdar_11}. 
In addition, we only discuss the case with the whole GNR under the ac gate
voltage, i.e., $N_b=0$. Thus for the total transmission 
$T_{LR/RL}(\varepsilon)=\sum_{\sigma n}T_{LR\sigma/RL\sigma}^n(\varepsilon)$, 
$T_{LR}(\varepsilon)=T_{RL}(\varepsilon)$ due to the symmetry of this system.
The other parameters are set to be $N_W=80$, $N_{ac}=400$ and $V_g=0$.

\begin{figure}[tbp]
  \begin{center}
    \includegraphics[width=6.1cm]{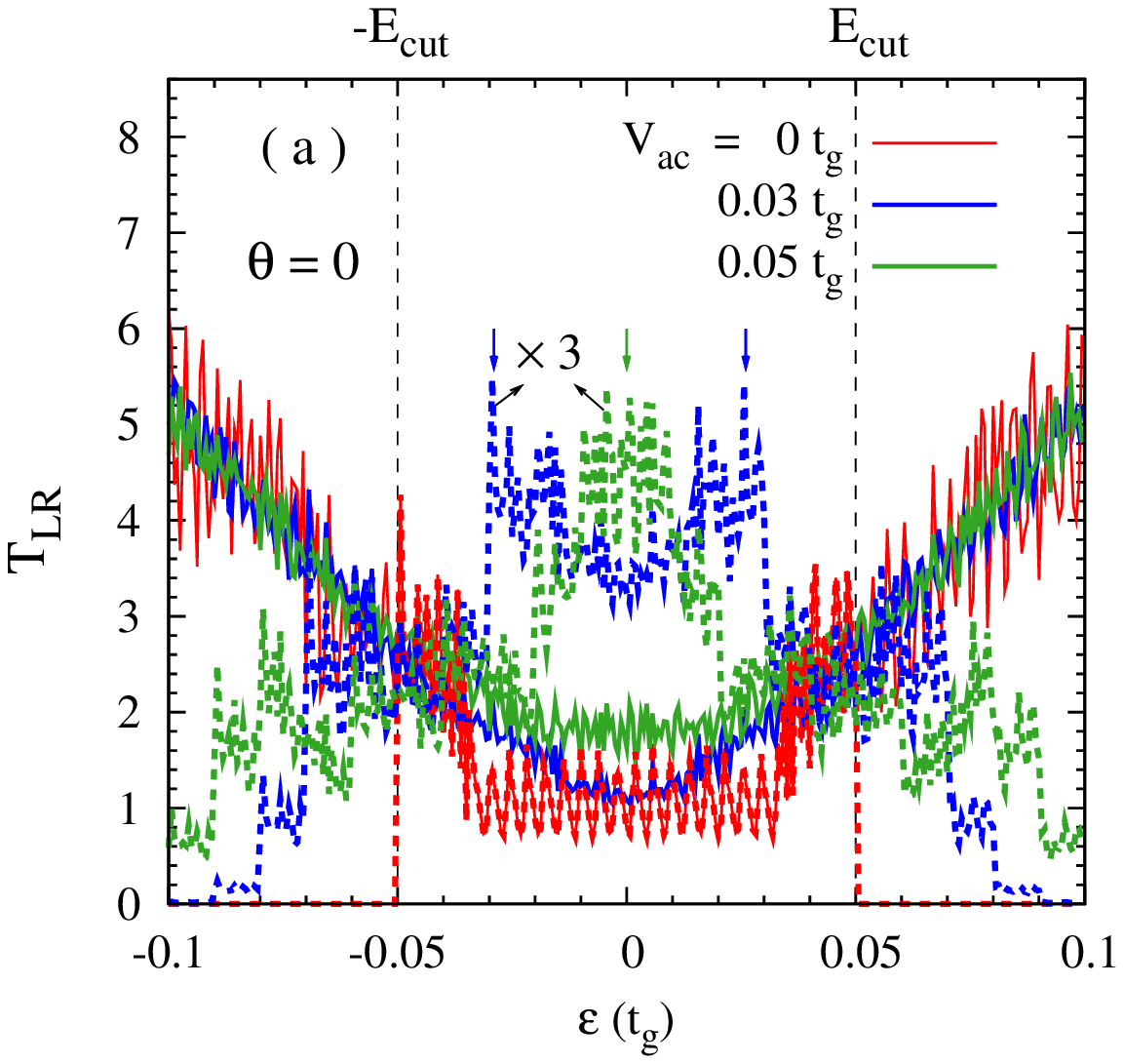}
    \includegraphics[width=6.1cm]{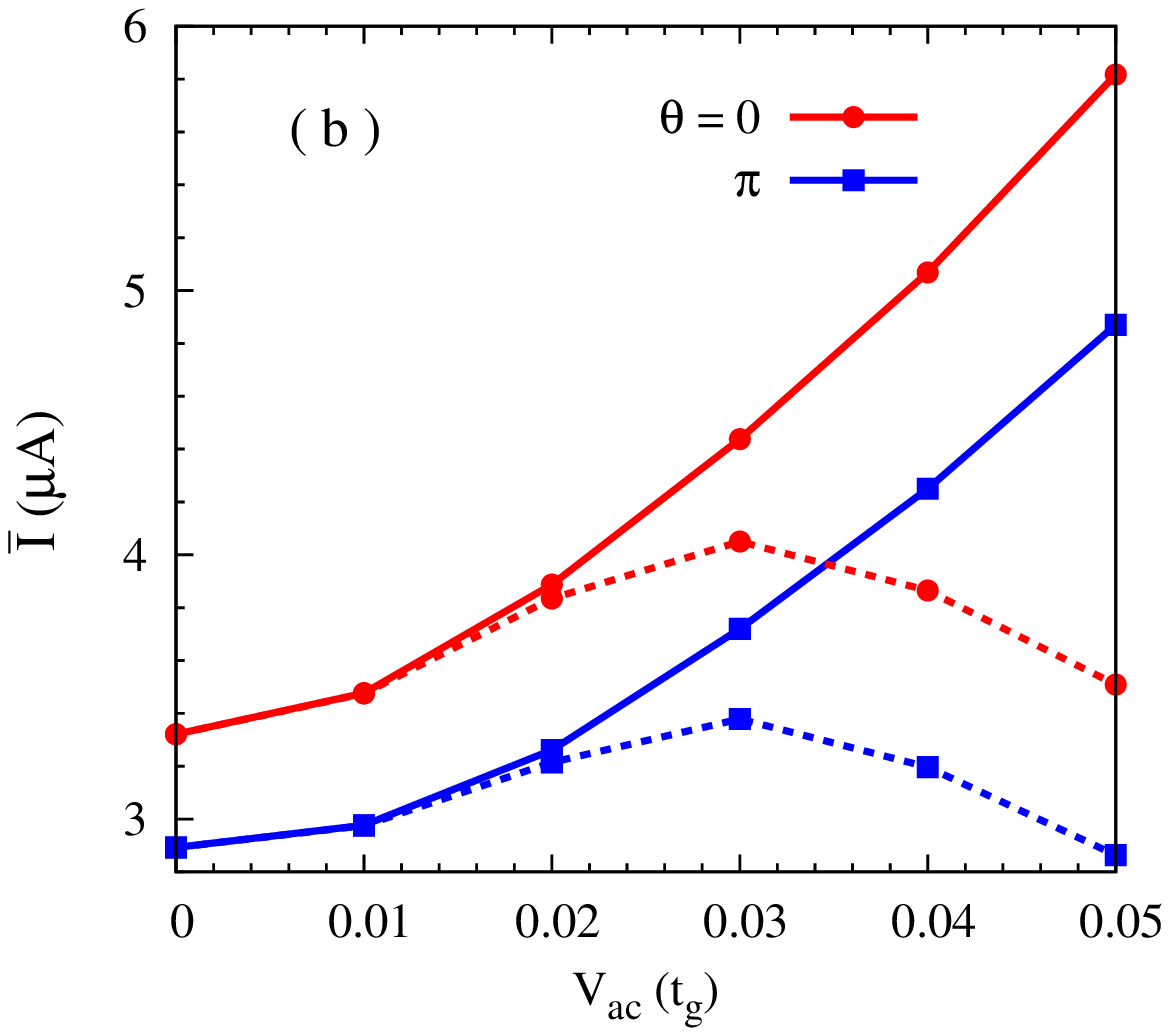} 
  \end{center}
  \caption{ (Color online) Armchair GNR connected with ferromagnetic leads.
    (a) Total transmission $T_{LR}(\varepsilon)$ as function of the energy
    $\varepsilon$ with different ac-field strength $V_{ac}$ in the parallel
    configuration of the lead magnetizations. 
    (b) Time-averaged currents in the parallel and antiparallel configurations
    as function of the ac-field strength $V_{ac}$.
    $\Omega=0.01 t_g$ in these figures. The dashed curves represent the results
    with the cutoff energy $E_{\rm cut}=0.05 t_g$. 
    The blue and green arrows in (a) indicate the positions of the pronounced
    peaks in the blue and green dashed curves, respectively. 
  }
  \label{fig_com} 
\end{figure}

We first compare the total transmission $T_{LR}(\varepsilon)$ with and without
the introduction of the cutoff energy. 
In Fig.~\ref{fig_com}(a), we plot $T_{LR}(\varepsilon)$ as function of the
energy $\varepsilon$ with different ac-field strengths $V_{ac}$ for the field 
frequency $\Omega=0.01 t_g$ in the parallel configuration of the lead
magnetizations.\cite{similar} 
Here the solid curves represent the results without a cutoff, while the
dash ones represent the results from the calculations by replacing
Eq.~(\ref{G_0_2}) with 
\begin{eqnarray}
  \nonumber
  \overline{g}^r(\varepsilon_r,n,m)&=&\sum_N J_{n-N}(\frac{V_{ac}}{\Omega})
  J_{m-N}(\frac{V_{ac}}{\Omega}) \hat{g}^r_0(\varepsilon_r+N\Omega) \\
  &&{}\times \Theta(E_{\rm cut}-|\varepsilon_r+N\Omega|)
  \label{G_0_cut}
\end{eqnarray}
with the cutoff energy $E_{\rm cut}=0.05 t_g$.
Our results show that after introducing the cutoff energy in such a way, the
field-free transmission is the same as the one without a cutoff when
$|\varepsilon|<E_{\rm cut}$, but becomes zero for $|\varepsilon|>E_{\rm cut}$. 

The difference between the transmissions with and without  
the cutoff energy becomes more pronounced in the presence of the ac field.
In particular, some pronounced peaks appear in the transmissions with the
cutoff energy as indicated by the green and blue arrows,
which should be absent if calculated properly. 
This phenomenon can be understood via the Tien-Gordon
theory,\cite{Torres_ac_trans,Tien_63,Hanggi_98,Hanggi_05} which is valid for a
weak energy dependence of the self-energy from the leads. 
This theory gives the formula
\begin{equation}
  T_{LR}(\varepsilon)=
  \sum_{m} \left[J_{m}(\frac{V_{ac}}{\Omega})\right]^2 
  T_0(\varepsilon+m\Omega),
  \label{T_flatband}
\end{equation} 
where $T_0(\varepsilon)= {\rm Tr}\left\{  \hat{\Gamma}_L(\varepsilon)
  \hat{G}^r_0(\varepsilon) \hat{\Gamma}_R(\varepsilon) \hat{G}^a_0(\varepsilon) 
\right\}$ is the field-free transmission.
The results from this formula are verified to coincide with those
from our exact calculations. 
This theory shows that the field-applied transmission is just the weighted
average of the field-free transmissions corresponding to various sidebands.
After the cutoff energy is introduced, $T_0(\varepsilon+m\Omega)$ in
Eq.~(\ref{T_flatband}) becomes zero for $|\varepsilon+m\Omega|>E_{\rm cut}$.
Thus, the corresponding sidebands cannot contribute to the field-applied
transmission. 
When $|\varepsilon|$ is small, the influence of the cutoff energy
is still weak, hence the transmission tends to increase with increasing
$|\varepsilon|$ just as the one without a cutoff.
Nevertheless, at large $|\varepsilon|$, the influence of the cutoff
energy becomes very important. 
In this case, the transmission tends to decrease 
with $|\varepsilon|$ due to the absence of the contribution of the sidebands 
satisfying $|\varepsilon+m\Omega|>E_{\rm cut}$. 
As a result, the pronounced peaks in the transmission are formed as indicated by
the blue arrows.  
Moreover, when the ac field is strong enough, the decreasing trend from the
cutoff energy is always dominant. Thus, the pronounced peak moves to the 
Dirac point as indicated by the green arrow.
Based on the above discussions, one can conclude that this kind of peaks,
which also appear in Fig.~2(a) in Ref.~\onlinecite{Berakdar_11}, are just the
artificial results caused by the introduction of the cutoff energy.

Similar problems can be found in the field-strength dependence of the
time-averaged current. We plot the time-averaged currents
$\overline{I}({\theta})$ in the parallel ($\theta=0$) and antiparallel
($\theta=\pi$) configurations against the field strength $V_{ac}$ for the Fermi
energies $E_{\rm F}^L=0.03 t_g$ and $E_{\rm F}^R=0$ in Fig.~\ref{fig_com}(b). 
It is shown that without a cutoff energy, the currents in both configurations
increase monotonically with increasing $V_{ac}$. 
This feature is from the increasing weight of the sidebands outside the energy
regime between the Fermi energies of two leads, where the field-free
transmissions are larger than those inside this energy regime as shown in
Fig.~\ref{fig_com}(a). 
However, after the introduction of the cutoff energy, the behaviour becomes very
different: the currents first increase and then decrease with increasing $V_{ac}$.
The decrease at high $V_{ac}$ is from the increasing weight of the sideband
beyond the cutoff energy, where field-free transmission is zero.
This means that this decrease, which also appears in Fig.~1(b) in
Ref.~\onlinecite{Berakdar_11}, is an artificial result as well. 



\begin{thebibliography}{0}
\bibitem{similar} The behaviours of the transmissions in the anti-parallel
configuration of the lead magnetizations are similar to that in the parallel
configuration, and hence not repeated here.
\end{thebibliography}
\end{document}